\documentclass[12pt]{article}

\usepackage[pdftex]{graphicx}
\usepackage{epstopdf}
\usepackage{natbib}
\usepackage{geometry,rotate,lscape,epsfig,booktabs,
	colortbl,xspace,sectsty,amsthm,float,xy,xr,enumitem,
	amssymb,amsfonts,amsbsy,amsmath,mathtools,delarray,mathtools,
	bm,color,multirow,latexsym,fancyhdr,fontenc,anysize,url,caption,subcaption,
	rotating,ifthen,fancybox,fullpage,amscd,pifont,setspace,xargs}
\usepackage[mathscr]{eucal} 
\usepackage{cleveref}
\usepackage{wrapfig}

\usepackage{titlesec}
\titleformat*{\section}{\large\bfseries\sffamily}
\titleformat*{\subsection}{\small\bfseries\sffamily}
\titleformat*{\subsubsection}{\footnotesize\bfseries\sffamily}
\bibpunct{(}{)}{;}{a}{}{,}
\crefformat{equation}{(#2#1#3)}
\crefrangeformat{equation}{(#3#1#4) to~(#5#2#6)}
\crefname{equation}{}{}
\Crefname{Equation}{}{}

\definecolor{DarkBlue}{rgb}{0,.08,.45}
\definecolor{DarkRed}{rgb}{.7,0,.4}

\def\hg #1 {\texcolor{cyan}{{\it Hans:}   #1}}

\newcommand{\be}{\begin{equation}}
\newcommand{\ee}{\end{equation}}
\newcommand{\ben}{\begin{equation}}
\newcommand{\een}{\end{equation}}
\newcommand{\ba}{\begin{align}}
\newcommand{\ea}{\end{align}}
\newcommand{\ban}{\begin{align*}}
\newcommand{\ean}{\end{align*}}
\newcommand{\bsp}{\begin{split}}
\newcommand{\esp}{\end{split}}
\newcommand{\ed}{\end{document}}

\newcommand{\btab}{\begin{tabular}}
\newcommand{\etab}{\end{tabular}}
\newcommand{\bc}{\begin{center}}
\newcommand{\ec}{\end{center}}

\newcommand{\bi}{\begin{itemize}}
\newcommand{\ei}{\end{itemize}}
\newcommand{\bfi}{\begin{figure}}
\newcommand{\efi}{\end{figure}}
\newcommand{\bem}{\begin{enumerate}}
\newcommand{\eem}{\end{enumerate}}
\newcommand{\bdes}{\begin{description}}
\newcommand{\edes}{\end{description}}
\newcommand{\bay}{\begin{array}}
\newcommand{\eay}{\end{array}}
\newcommand{\bs}{\boldsymbol}

\newtheorem{thm}{Theorem}
\newtheorem{lem}{Lemma}

\newcommand{\single}{\renewcommand{\baselinestretch}{1.2}\normalsize}
\newcommand{\double}{\renewcommand{\baselinestretch}{1.6}\normalsize}

\def\bco{\iffalse}

\def\Var{{\rm Var}}

\def\f{Fr\'echet }

\def\i1{_{i1}}

\DeclareMathOperator*{\argmin}{argmin}
\def\argmin#1{\mathrel{\mathop{\arg\min}\limits_{#1}}}

\def\cW{\mathcal{W}}

\newcommand{\bpmt}{\begin{pmatrix}}
\newcommand{\epmt}{\end{pmatrix}}

\def\mc{\mathcal}

\def\mr{\mathrm}

\def\s1n{\sum_{i=1}^n}
\def\p1n{\prod_{i=1}^n}
\def\i01{\int_0^1}

\usepackage{tabularx}
\usepackage{longtable}
\usepackage{tikz-cd}
\usepackage[linesnumbered,ruled,vlined]{algorithm2e}

\def\T{{ \mathrm{\scriptscriptstyle T} }}

\title{Wasserstein Regression with Empirical Measures and Density Estimation for Sparse Data}
\author{Yidong Zhou and Hans-Georg M\"{u}ller\\
	Department of Statistics, University of California, Davis\\
	Davis, CA 95616, USA}
\date{July 27, 2023}

\begin{document}
\maketitle

\begin{abstract}
	The problem of modeling the relationship between univariate distributions and one or more explanatory variables lately has found increasing interest. Traditional functional data methods cannot be applied directly to distributional data because of their inherent constraints. Modeling distributions as elements of the Wasserstein space, a geodesic metric space equipped with the Wasserstein metric that is related to optimal transport, is attractive for statistical applications. Existing approaches proceed by substituting proxy estimated distributions for the typically unknown response distributions. These estimates are obtained from available data but are problematic when for some of the distributions only few data are available. Such situations are common in practice and cannot be addressed with currently available approaches, especially when one aims at density estimates. We show how this and other problems associated with density estimation such as tuning parameter selection and bias issues can be side-stepped when covariates are available. We also introduce a novel version of distribution-response regression that is based on empirical measures. By avoiding the preprocessing step of recovering complete individual response distributions, the proposed approach is applicable when the sample size available for each distribution varies and  especially when it is small for some of the distributions but large for others. In this case, one can still obtain consistent distribution estimates even for distributions with only few data by gaining strength across the entire sample of distributions, while traditional approaches where distributions or densities are estimated individually fail, since sparsely sampled densities cannot be consistently estimated. The proposed model is demonstrated to outperform existing approaches through simulations. Its efficacy is corroborated in two case studies on Environmental Influences on Child Health Outcomes (ECHO) data and eBay auction data.
	
\end{abstract}

\double
\section{Introduction}
Data consisting of samples of univariate probability distributions are increasingly prevalent across various research areas. Data that contain distributions  as a basic  unit of observation are encountered in the analysis of mortality \citep{chen:23:1,ghod:21}, population pyramids \citep{hron:16, bigo:17}, brain connectivity \citep{mull:16:1}, and financial returns \citep{zhan:22, mull:23:2}, among many other applications. This has led to the emerging field of distributional data analysis \citep{pete:22}. Distributions, represented as probability density functions, cumulative distribution functions or quantile functions, come with inherent constraints that are not present in traditional functional data. The space where distributions are situated is not a vector space, and linear methods developed for functional data cannot be directly applied. Various approaches have been proposed  to address this challenge, including  global transformations of univariate distributions to a Hilbert space \citep{hron:16, mull:16:1}, which however do not take  the geometry of the space of distributions  into account and are not isometric.  

More recently, attention has focused on directly modeling distributions as elements of the Wasserstein space, a geodesic metric space related to optimal transport \citep{vill:03,pana:20}. The pseudo-Riemannian structure of this space can be used to construct isometric exponential maps from the space to tangent bundles, where linear operations can be deployed \citep{bigo:17, chen:23:1}; however the inverse log map is not defined on the entire tangent space, which causes substantial difficulties  and requires an ad-hoc constraint or projection step \citep{flet:13, pego:22}. 

A  commonly encountered problem is to model the relationship between distributions and one or more explanatory variables. Distribution-response regression problems arise in many modern data analysis settings. Examples include neuroimaging, where varying patterns of brain connectivity distributions across age and other clinical covariates are of interest \citep{mull:16:1}, metabolomics, where the aim is to model the dependence of a  metabolite distribution on birth weight \citep{tals:18}, and also mortality where one is interested in the dependence of age-at-death distributions on economic indicators \citep{pete:22}. Using the Wasserstein metric, distribution-response regression can be implemented as a special case of \f regression \citep{mull:19:3}, which models the relationship between random objects that lie in a generic metric space as responses and scalar or vector covariates as predictors. 

In line with all local and global transformation methods, \f regression  requires that each distribution is observed as a distributional object. This is however  usually unrealistic in practice as one rarely  observes data samples where the atoms of the samples are entire distributions. Instead, one usually has samples of independent data $\{Y_{ij}\}_{j=1}^{N_i}$ that are generated by the distribution $\nu_i$ for each $i$; see Figure~\ref{fig:diagram} in Section \ref{subsec:imp} for a demonstration. To address this, current approaches for virtually all distributional data analysis methods include a prior distribution estimation step where one substitutes a smooth estimate for the  unobservable distributional objects. Commonly  this is done through a density estimate such as a kernel estimate or smoothed histogram that is obtained from data $\{Y_{ij}\}_{j=1}^{N_i}$ \citep{pana:16, bigo:18, mull:19:3} or a quantile function obtained by smoothing the empirical quantile function \citep{pete:21, mull:21:10}. 

In the current literature, 
strong assumptions are used to ensure uniform convergence of density estimates, uniform across all random distributions \citep{pana:16,mull:16:1, pete:21, mull:21:10}. This preliminary smoothing step  
requires the minimum number of observations $m=m(n)$ that are generated by each random distribution, where $n$ is the number of random distributions, to increase to infinity at a fast rate, typically faster than $n$. This entails convergence of the density estimation step at a rate that is faster than that of the subsequent \f regression or other operation so that the preliminary estimation step can be ignored in the overall asymptotic analysis. However, the assumption of a fast increase in the number of observations made for each of the distributions across the board is often unrealistic and the need to choose a tuning parameter for each distribution is another downside.

This state of affairs motivates the approach proposed in this paper, namely to use  empirical measures $\hat{\nu}_i=(1/N_i)\sum_{j=1}^{N_i}\delta_{Y_{ij}}$, where $N_i \ge 1$ and $\delta_{Y_{ij}}$ denotes the Dirac measure at $Y_{ij}$, i.e., using the observations made for each random distribution directly, rather than first forming density estimates or other distributional estimates when conducting \f regression. Besides avoiding the preliminary distribution estimation step with all its downsides, we will show that this has the further major benefit that it leads to a consistent density estimate for a distribution for which one has only very few observations, where the number of observations may not even increase with $n$ and density estimation would normally be a hopeless enterprise. This becomes possible by harnessing the data across all distributions and exploiting the assumed smooth dependency of the response distributions on the predictors.

The proposed regression model, implemented through least common multiples and supported with asymptotic theory, avoids  smoothing bias and tuning parameter choice in the pre-smoothing step and hence is more broadly applicable in practice, especially when the available sample size for each distribution greatly varies \citep{qiu:22:2}. A typical  example is provided by multi-cohort studies \citep{ocon:22}, where the availability and cost of observations highly varies across cohorts \citep{bone:14}. We illustrate such a scenario with Environmental influences on Child Health Outcomes (ECHO) data \citep{knap:23}, where the dependence of body mass index distributions for children across cohorts on demographic covariates is of interest. A second example is provided by point processes, where distributions of event arrival times are inherently discretely observed with heterogeneous sample sizes. This scenario is demonstrated with eBay auction data \citep{jank:10}, where we investigate the relationship between the bid time distribution and the opening bid.

The main innovations and strengths of the proposed approach are as follows. First, we illustrate that the universally employed pre-smoothing step is superfluous for distribution-response \f regression; omitting this step avoids initial smoothing bias and tuning parameter selection and also avoids the requirement of a rapidly increasing number of observations per distribution. Second, we obtain consistent density estimates for densities with sparse number of observations by borrowing strength across the sample. Third, the proposed regression model is shown to outperform the existing smoothing approach in finite sample situations. Fourth, the proposed methods are supported by theory, including pointwise and uniform rates of convergence. Fifth, we demonstrate the utility and flexibility of the proposed distribution-response regression model.

The rest of the paper is organized as follows. In Section \ref{sec:pre}, we introduce some basic background and notations for the Wasserstein space of distributions.  The proposed regression model for empirical measure responses and vector covariates is introduced in Section \ref{sec:emr}. Pointwise and uniform rates of convergence for the estimators are established in Section \ref{sec:th}. Computational details and simulation results are presented in Section 5. The proposed framework is illustrated in Section 6 using the ECHO and eBay auction data. 
Detailed theoretical proofs are in the Appendix.

\section{Preliminaries}
\label{sec:pre}
For a closed interval $\Omega$ 
with Borel $\sigma$-algebra $\mathcal{B}(\Omega)$, we denote the set of probability measures, also referred to as measures or distributions,  $\mu$ with domain $\Omega$ as  $\mathcal{P}(\Omega)$ and  define the space of measures with finite second moments as   
\begin{equation*}
	\mathcal{W}:=\left\{\mu\in \mathcal{P}(\Omega); \int_\Omega x^2\mu(dx)<\infty\right\}.
\end{equation*}
For any measure $\mu\in\mathcal{W}$ with cumulative distribution function $F_{\mu}$, we consider the quantile function $F_{\mu}^{-1}$ to be the left continuous inverse of $F_{\mu}$, i.e., $F_{\mu}^{-1}(\alpha)=\inf\{x\in\Omega: F_{\mu}(x)\geq\alpha\}$, for $\alpha\in(0, 1)$. The space $\mathcal{W}$ is a metric space with the 2-Wasserstein, or simply Wasserstein distance between two measures $\mu_1, \mu_2\in\mathcal{W}$ defined as
\[d_\mathcal{W}^2(\mu_1, \mu_2):=\inf_{\pi\in\Pi(\mu_1, \mu_2)}\int_{\Omega\times\Omega}|x-y|^2d\pi(x, y),\]
where $\Pi(\mu_1, \mu_2)$ is the set of joint measures on $\Omega\times\Omega$ with marginals $\mu_1$ and $\mu_2$ \citep{kant:42}.  It is well known 
\citep{vill:03} that the Wasserstein distance can be expressed as the $L^2$ distance between the corresponding quantile functions,
\begin{equation}
	\label{eq:dw}
	d_\mathcal{W}^2(\mu_1, \mu_2)=\int_0^1\{F_{\mu_1}^{-1}(\alpha)-F_{\mu_2}^{-1}(\alpha)\}^2d\alpha.
\end{equation}
It can be shown that $\mathcal{W}$ endowed with $d_{\mathcal{W}}$ is a complete and separable metric space, the Wasserstein space  \citep{pana:20, vill:03}. 

Consider a random element $\nu$ taking values in the Wasserstein space $\mathcal{W}$, assumed to be square integrable in the sense that $E\{d_{\mathcal{W}}^2(\nu, \mu)\}<\infty$ for 
all $\mu\in\mathcal{W}$. The \f mean of $\nu$ \citep{frec:48},  extending the usual notion of mean, is 
\[\nu_{\oplus}=\argmin{\mu\in\mathcal{W}}E\{d_{\mathcal{W}}^2(\nu, \mu)\},\]
which is well-defined and unique as the Wasserstein space $\mathcal{W}$ is a Hadamard space \citep{kloe:10}. It follows from \eqref{eq:dw} that the quantile function of the \f mean $\nu_{\oplus}$ is
$F^{-1}_{\nu_{\oplus}}(\cdot)=E\{F_{\nu}^{-1}(\cdot)\}.$
In the following, the notation $\mu$ refers to fixed measures in $\mathcal{W}$, $\nu$ to random measures, $a\lesssim b$ means that  there exists a positive constant $C$ such that $a\leq Cb$ and  $a\asymp b$ that $a\lesssim b$ and $b\lesssim a$. The Euclidean norm in $\mathbb{R}^p$ is denoted by $\|\cdot\|_E$.

\section{Wasserstein Regression with Empirical Measures}
\label{sec:emr}
Let $(Z, \nu)$ be a random pair with joint distribution $\mathcal{F}$ on the product space $\mathbb{R}^p\times\mathcal{W}$. Denote the mean and variance of $Z$ by $\theta=E(Z)$ and $\Sigma=\Var(Z)$, with $\Sigma$ strictly positive definite. To model the regression relation between random measures $\nu$ and vector covariates $Z$, we adopt the framework of \f regression, a version of conditional \f means designed for the regression of metric space-valued responses on Euclidean predictors \citep{mull:19:3, mull:22:8}. \f regression  targets the conditional \f mean of $\nu$ given $Z=z$, 
\begin{equation*}
    m(z)=\argmin{\mu\in\mathcal{W}}E\{d_{\mathcal{W}}^2(\nu, \mu)|Z=z\}.
\end{equation*}
A brief description of \f regression is as follows; for more details see \citet{mull:19:3}. Suppose that $\{(Z_i, \nu_i)\}_{i=1}^n$ are $n$ independent realizations of $(Z, \nu)$ with 
\[\bar{Z}=\frac{1}{n}\sum_{i=1}^nZ_i,\quad \hat{\Sigma}=\frac{1}{n}\sum_{i=1}^n(Z_i-\bar{Z})(Z_i-\bar{Z})^\T.\]

Global \f regression can be considered as a  generalization of classical multiple linear regression for real-valued responses and targets 
\begin{equation}
    \label{eq:mg}
    m_G(z)=\argmin{\mu\in\mathcal{W}}E\{s_G(z)d_{\mathcal{W}}^2(\nu, \mu)\},
\end{equation}
with weight function $s_G(z)=1+(Z-\theta)^\T\Sigma^{-1}(z-\theta)$ and empirical version
\begin{equation}
    \label{eq:hatmg}
    \tilde{m}_G(z)=\argmin{\mu\in\mathcal{W}}\frac{1}{n}\sum_{i=1}^ns_{iG}(z)d^2_{\mathcal{W}}(\nu_i, \mu),
\end{equation}
 for a sample $\{(Z_i, \nu_i)\}_{i=1}^n$ , where $s_{iG}(z)=1+(Z_i-\bar{Z})^\T\hat{\Sigma}^{-1}(z-\bar{Z})$. 

Analogously, local \f regression extends local linear regression to metric space-valued responses. For the special case of a scalar predictor $Z\in\mathbb{R}$ it targets 
\begin{equation}
    \label{eq:ml}
    m_{L, h}(z)=\argmin{\mu\in\mathcal{W}}E\{s_{L}(z, h)d_{\mathcal{W}}^2(\nu, \mu)\},
\end{equation}
where $s_L(z, h)=K_h(Z-z)[u_2-u_1(Z-z)]/\sigma_0^2$,  $u_j=E[K_h(Z-z)(Z-z)^j]$ for $j=0, 1, 2$,  $\sigma_0^2=u_0u_2-u_1^2$, and $K_h(\cdot)=h^{-1}K(\cdot/h)$ with $K(\cdot)$ a continuous symmetric probability density function on $[-1, 1]$ and $h$ a bandwidth and with empirical version
\begin{equation}
    \label{eq:hatml}
    \tilde{m}_{L, h}(z)=\argmin{\mu\in\mathcal{W}}\frac{1}{n}\sum_{i=1}^ns_{iL}(z, h)d_{\mathcal{W}}^2(\nu_i, \mu).
\end{equation}
Here $s_{iL}(z, h)=K_h(Z_i-z)[\hat{u}_2-\hat{u}_1(Z_i-z)]/\hat{\sigma}_0^2$, where $\hat{u}_j=n^{-1}\sum_{i=1}^nK_h(Z_i-z)(Z_i-z)^j$ for $j=0, 1, 2$ and $\hat{\sigma}_0^2=\hat{u}_0\hat{u}_2-\hat{u}_1^2$.

In previous work on Fr\'echet regression \citep{mull:19:3}, response distributions in the Wasserstein space $\mathcal{W}$ served as one of the key examples but it has been typically assumed that the measures are fully observed while at the same time they may be randomly perturbed in analogy to usual additive noise models \citep{mull:22:8}. However none of this applies in the more realistic situation where one has data $\{Y_{ij}\}_{j=1}^{N_i}$ that are generated from each distribution $\nu_i$. The stopgap solution applied previously is a preprocessing density estimation step \citep{pana:16,mull:16:1}, where it is assumed that the number of data available for all measures increases at a faster rate than the available number of observation points $(Z_i,\nu_i)$ in order to preserve asymptotic convergence rates. 

However the intermediate kernel density estimates will be biased and inconsistent if some of the measures generate only very few observations, which then renders this approach infeasible. This 
provides the motivation to replace the unobservable measures $\nu_i$ with the empirical measure $\hat{\nu}_i=(1/{N_i})\sum_{j=1}^{N_i}\delta_{Y_{ij}}$. This circumvents the pre-smoothing step and hence eliminates the corresponding tuning parameter selection and smoothing bias. Using empirical measures $\hat{\nu}_i$ in lieu of the unobservable measures $\nu_i$ as responses, the approaches proposed here are the global and local Regression with Empirical Measures (REM), replacing  \eqref{eq:hatmg} and \eqref{eq:hatml} by 
\begin{align}
	&\hat{m}_G(z)=\argmin{\mu\in\mathcal{W}}\frac{1}{n}\sum_{i=1}^ns_{iG}(z)d^2_{\mathcal{W}}(\hat{\nu}_i, \mu),\label{eq:tildemg}\\&
	\hat{m}_{L, h}(z)=\argmin{\mu\in\mathcal{W}}\frac{1}{n}\sum_{i=1}^ns_{iL}(z, h)d_{\mathcal{W}}^2(\hat{\nu}_i, \mu).\label{eq:tildeml}
\end{align}
The computation of the minimizers in \eqref{eq:tildemg} and \eqref{eq:tildeml} is not straightforward, as some of the weights in these minimization problems are inherently negative. We show in Section \ref{subsec:imp} that  adopting least common multiples leads to a simple and efficient algorithm. 

\section{Asymptotic Properties}
\label{sec:th}
We establish pointwise and uniform rates of convergence under the framework of M-estimation. Denote by $\mathcal{W}^{\text{ac}}$ the set of measures $\mu\in\mathcal{W}$ that are absolutely continuous with respect to the Lebesgue measure $dx$ on $\mathbb{R}$. To establish rates of convergence for  estimates \eqref{eq:tildemg} and \eqref{eq:tildeml}, we require the following conditions.
\bem[label=(C\arabic*), leftmargin=1cm]
\item The random measure $\nu$ satisfies $\nu\in\mathcal{W}^{\text{ac}}$ and its probability density function $f_\nu$ is bounded below by a positive constant.\label{itm:A1}
\item The sample size $N_i$ is independently Poisson distributed with parameter $c_i\lambda_n$ where $c_i\geq c>0$ are  constants  and $0<\lambda_n/\log n\to\infty$ as $n\to\infty$. \label{itm:A2}
\eem
Condition \ref{itm:A1} ensures the random measure $\nu$ is well-behaved. 
Since measures $\nu_i$ are discretely observed,  rates of convergence will be affected by the rate at which the sample size $N_i$ increases to infinity. Instead of directly requiring the sample size $N_i$ to diverge to infinity, Condition \ref{itm:A2} imposes a distributional condition on $N_i$. 
This set-up reflects the case of heterogeneous sample sizes, as it allows some sample sizes $N_i$ to be very small and even $0$ for some of the observed distributions, but also 
ensures that the $N_i$ are positive almost surely for large enough $n$; compare  Lemma 3 in \citet{pana:16}. The following result formalizes the consistency of the proposed global REM estimates and provides rates of convergence.

\begin{thm}
\label{thm:1}
Under Conditions \ref{itm:A1} and \ref{itm:A2}, the global REM estimate defined in \eqref{eq:tildemg} satisfies
\begin{equation*}
	d_\mathcal{W}\{\hat{m}_G(z), m_G(z)\}=O_p(n^{-1/2}+\lambda_n^{-1/2}).
\end{equation*}
Furthermore, for any constant  $B$ it holds that for any $\varepsilon>0$, 
\[\sup_{\|z\|_E\leq B}d_{\mathcal{W}}\{\hat{m}_G(z), m_G(z)\}=O_p(n^{-1/\{2(1+\varepsilon)\}}+\lambda_n^{-1/2}).\]
\end{thm}

All proofs are in the Appendix. The pointwise rate of convergence is $O_p(n^{-1/2})$ as long as the sample size $N_i$, on average, grows at the same rate as $n$,  i.e., $\lambda_n\asymp n$, which  is the same as the well-known optimal rate for multiple linear regression. Similarly we obtain the following result for  local REM, where the  kernel and distributional assumptions \ref{itm:lp0}--\ref{itm:lu1} listed in the Appendix A are standard for local regression.

\begin{thm}
\label{thm:2}
Under Conditions \ref{itm:A1} and \ref{itm:A2}, the local REM estimate defined in \eqref{eq:tildeml} satisfies
\begin{equation*}
	d_\mathcal{W}\{\hat{m}_{L, h}(z), m(z)\}=O_p(n^{-2/5}+\lambda_n^{-1/2})
\end{equation*}
for $h\sim n^{-1/5}$ if Assupmtions \ref{itm:lp0} and \ref{itm:lp1} hold. Furthermore, for a closed interval $\mathcal{T}\subset\mathbb{R}$, 
\[\sup_{z\in\mathcal{T}}d_\mathcal{W}\{\hat{m}_{L, h}(z), m(z)\}=O_p(n^{-1/(3+\varepsilon)}+\lambda_n^{-1/2})\]
for $h\sim n^{-1/(6+2\varepsilon)}$ and any $\varepsilon>0$ if Assumptions \ref{itm:lu0} and \ref{itm:lu1} hold.
\end{thm}

As long as $\lambda_n\asymp n^{4/5}$, the local REM estimate achieves the pointwise rate $O_p(n^{-2/5})$, corresponding to the well-known optimal rate for standard local linear regression. 

\section{Implementation and Simulations}
\subsection{Implementation details}
\label{subsec:imp}
To implement the proposed methods, one needs to solve minimization problems \eqref{eq:tildemg} and \eqref{eq:tildeml}. By standard properties of the $L^2(0, 1)$ inner product, \eqref{eq:tildemg} and \eqref{eq:tildeml} can be simplified to
\begin{align}
	&\hat{m}_G(z)=\argmin{\mu\in \mathcal{W}}d_{L^2}^2\{\hat{B}_G(z), F_{\mu}^{-1}\},\label{eq:bg}\\
	&\hat{m}_{L, h}(z)=\argmin{\mu\in \mathcal{W}}d_{L^2}^2\{\hat{B}_{L, h}(z), F_{\mu}^{-1}\},\nonumber
\end{align}
where $\hat{B}_G(z)=n^{-1}\sum_{i=1}^ns_{iG}(z)F_{\hat{\nu}_i}^{-1}$, $\hat{B}_{L, h}(z)=n^{-1}\sum_{i=1}^ns_{iL}(z, h)F_{\hat{\nu}_i}^{-1}$ and $d_{L^2}(\cdot, \cdot)$ denotes the $L^2$ distance; see the proof of Theorem \ref{thm:1} for details. The minimizers $\hat{m}_G(z)$ and $\hat{m}_{L, h}(z)$, viewed as projections onto $\mathcal{W}$, exist and are unique for any $z$ by convexity and closedness of $\mathcal{W}$. To tackle the challenge of substantial variation in the number of jumps for empirical quantile functions $F_{\hat{\nu}_i}^{-1}$ due to varying sample sizes $N_i$, we propose an algorithm based on the least common multiple, which allows us to compute $\hat{B}_G(z)$ and $\hat{B}_{L, h}(z)$ as straightforward weighted averages. The algorithm for global REM is outlined in Algorithm \ref{alg:lcm}. The only difference for local REM is the substitution of the weight function $s_{iL}(z, h)$ for $s_{iG}(z)$.

\begin{algorithm}[tb]
	\SetAlgoLined
	\KwIn{data $\{(Z_i, \{Y_{ij}\}_{j=1}^{N_i})\}_{i=1}^n$, and a new predictor level $z$.}
	\KwOut{prediction $\hat{m}_G(z)$.}
	$M\longleftarrow$ the least common multiple of $\{N_i\}_{i=1}^n$\;
	$\{\mathbb{V}_i\}_{i=1}^n\longleftarrow$ for each $i=1, \ldots, n$, stretch $(Y_{i1}, \ldots, Y_{iN_i})^\T$ to a vector of length $M$, $\mathbb{V}_i=(V_{i1}, \ldots, V_{iM})^\T$, by repeating each element $\frac{M}{N_i}$ times and then arranging in ascending order\;
	$s_{iG}(z)\longleftarrow1+(Z_i-\bar{Z})^\T\hat{\Sigma}^{-1}(z-\bar{Z})$ where $\bar{Z}=\frac{1}{n}\sum_{i=1}^n Z_i$ and $\hat{\Sigma}=\frac{1}{n}\sum_{i=1}^n(Z_i-\bar{Z})(Z_i-\bar{Z})^\T$ are the sample mean and variance of $\{Z_i\}_{i=1}^n$, respectively\;
	$\bar{\mathbb{V}}=(\bar{V}_1, \ldots, \bar{V}_M)^\T\longleftarrow\frac{1}{n}\sum_{i=1}^ns_{iG}(z)\mathbb{V}_i$, the element-wise weighted average of $\{\mathbb{V}_i\}_{i=1}^n$\;
	$\hat{m}_G(z)\longleftarrow\argmin{\mu\in\mathcal{W}}d_{L^2}^2(\hat{B}_G(z), F_{\mu}^{-1})$, where $\hat{B}_G(z)=\sum_{m=1}^M\bar{V}_m\mathbf{1}_{q\in(\frac{m-1}{M}, \frac{m}{M}]}$ is a step function defined on $q\in[0, 1]$.
	\caption{Global REM}
	\label{alg:lcm}
\end{algorithm}

In the first step of Algorithm \ref{alg:lcm}, one needs to calculate $M=\mathrm{lcm}(\{N_i\}_{i=1}^n)$,  the least common multiple of the ensemble $\{N_i\}_{i=1}^n$. 
To control the size of $\mathrm{lcm}(\{N_i\}_{i=1}^n)$  especially when $n$ is large and $N_i$ greatly varies, we incorporate a threshold $M=\min\{\mathrm{lcm}(\{N_i\}_{i=1}^n), M_0\}$, where $M_0$ is a pre-specified constant, say $5,000$. The fifth step of Algorithm \ref{alg:lcm} involves a minimization problem to obtain the projection onto $\mathcal{W}$. For this step, we  use a Riemann sum approximation of the $L^2$ distance. This leads to the following  convex quadratic optimization problem,
	\begin{align}
		\begin{split}
			&\text{minimize}\quad \sum_{m=1}^M(q_m-\bar{V}_m)^2\\
			&\text{subject to}\quad q_1\leq q_2\leq\cdots\leq q_{M-1}\leq q_M,\\
			&\hspace{0.87in}q_j\in\Omega,\quad j=1, \ldots, M.
		\end{split}
		\label{eq:osqp}
	\end{align}
The solution $q^*$ represents a discretized version of the predicted quantile function. We use the \texttt{osqp} package \citep{stel:20} in R to solve the optimization problem presented in \eqref{eq:osqp}.

\begin{figure}[tb]
	\centering
	\includegraphics[width=0.9\linewidth]{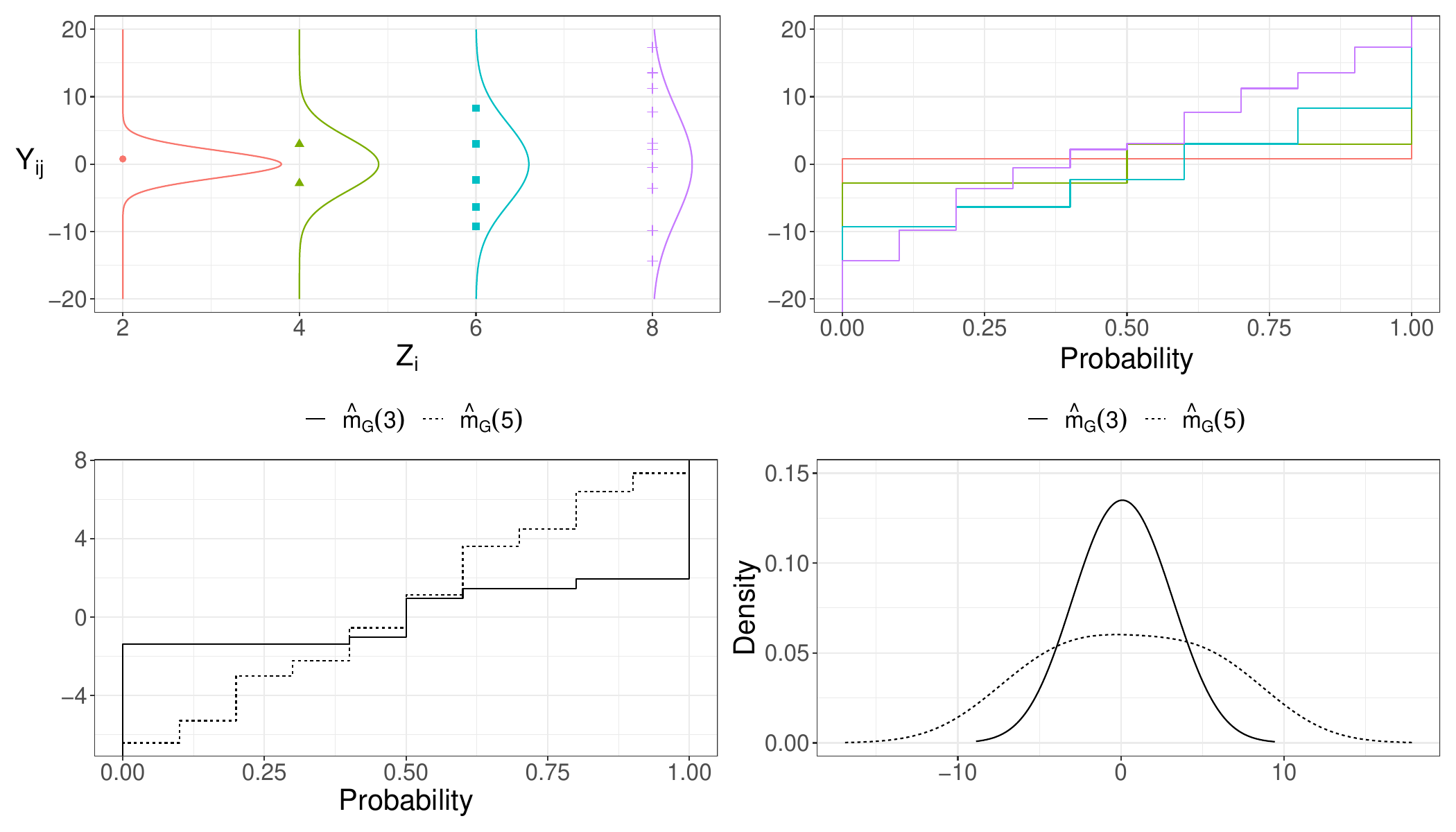}
	\caption{Illustrating global REM, where $N_i$ observations $\{Y_{ij}\}_{j=1}^{N_i}$ are sampled from distributions $\nu_i=N(0, Z_i^2)$ for $i=1, 2, 3, 4$, with sample sizes $N_1=1, N_2=2, N_3=5, N_4=10$ and scalar covariates $Z_1=2, Z_2=4, Z_3=6, Z_4=8$. Top left: Visualization of $Y_{ij}$ versus covariate levels $Z_i$, along with the underlying true densities. Top right: Empirical quantile functions corresponding to the empirical measures $\hat{\nu}_i=(1/{N_i})\sum_{j=1}^{N_i}\delta_{Y_{ij}}$. Bottom: Predicted quantile functions (left) and corresponding  densities (right), obtained with global REM for predictor levels $z=3$ and $z=5$.} 
	\label{fig:diagram}
\end{figure}

An illustrative example is in Figure~\ref{fig:diagram} to demonstrate  the proposed approach and algorithm for the case of global REM. In the top left panel of Figure~\ref{fig:diagram}, we depict large variation in sample sizes $N_i$ of the observations available for distributions $\nu_i$, where the smallest $N_i$ is $N_i=1.$  A simple calculation shows that the global weight function is $s_{iG}(z)=1+(2i-5)(z-5)/5$ for $i=1, 2, 3, 4$. The predicted distribution at predictor level $z$ is obtained by calculating $\hat{B}_G(z)$ in \eqref{eq:bg}, a weighted average of empirical quantile functions $F_{\hat{\nu}_i}^{-1}$, see the top right panel of Figure~\ref{fig:diagram}, utilizing weights $s_{iG}(z)$, followed by a projection onto $\mathcal{W}$. The weighted average is implemented  with Algorithm~\ref{alg:lcm}. The predicted distribution $\hat{m}_G(z)$ obtained in Algorithm~\ref{alg:lcm} is represented as an  empirical quantile function, which can subsequently be converted into a density function for improved visualization. We  adopted the qf2pdf function in the \texttt{frechet} package \citep{chen:20} to construct densities from quantile functions in this example and for our data applications; there are also alternative options  \citep{nile:22}.

The bottom two panels of Figure~\ref{fig:diagram} illustrate the predicted distributions using global REM for predictor levels $z=3$ and $z=5$. For $z=\bar{Z}=5$, one obtains the Wasserstein barycenter of the four empirical measures as for this case the global weight function is constant, i.e., $s_{iG}(5)=1$ for all $i$. Since $z=3$ is close to the left endpoint, the estimation requires negative weights where $s_{4G}(3)=-1/5$. In the presence of negative weights, $\hat{B}_G(3)$ no longer resides in $\mathcal{W}$, necessitating the additional projection step as per \eqref{eq:bg}.

\subsection{Simulations}
To assess the finite sample performance of the proposed methods, we construct a generative model that produces random responses $\nu$ along with an Euclidean predictor $Z\in\mathbb{R}$. Consider the true regression function $m(Z)$, represented as quantile function, 
\[m(Z)=E(\eta|Z)+E(\sigma|Z)\Phi^{-1}(\cdot),\]
which corresponds to a Gaussian distribution with mean and standard deviation depending on $Z$.
The distribution parameters of the true regression function $m(Z)$ are generated conditionally on $Z$, where the mean and the standard deviation are assumed to follow a normal distribution and a Gamma distribution, respectively. Four different simulation settings are examined as summarized in Table~\ref{tab:simusetting}. In Settings I and II, the response is generated, on average, as a Gaussian distribution with parameters depending on $Z$. Setting I corresponds to a global scenario where the response $\nu$ depends linearly on the predictor $Z$ as $E(\eta|Z)=\eta_0+\alpha Z$ and $E(\sigma|Z)=\sigma_0+\beta Z$. In Setting II, the nonlinear relationships 
$E(\eta|Z)=\eta_0+\alpha\sin(\pi Z)$ and $E(\sigma|Z)=\sigma_0+\beta\sin(\pi Z)$ is considered for the true underlying regression model. 

To take non-Gaussian distributions into consideration, we apply an additional transportation to the random response in Settings III and IV. Specifically, after sampling the distribution parameters as in the previous settings, the resulting distribution is transported in Wasserstein space via a random transport map $T$, uniformly sampled from the collection of maps $T_k(\alpha)=\alpha-\sin(k\alpha)/|k|$ for $k\in\{-2, -1, 1, 2\}$ \citep{pana:16}. Such a transportation significantly complicates Settings III and IV and makes the response distribution non-Gaussian. One can show that the true regression function remains the same after this random transportation.

\begin{table}[tb]
	\caption{Four simulation settings,  where $F_{\nu}^{-1}$ represents the quantile function of the generated random response. The distribution parameters of the random response depend on the predictor $Z$ as indicated for Settings I--IV. An additional transport map $T$, uniformly sampled from the collection of maps $T_k(\alpha)=\alpha-\sin(k\alpha)/|k|$ for $k\in\{-2, -1, 1, 2\}$, is applied to the resulting random responses in Settings III and IV.}
	\centering
	\btab{l|ll}
		\hline
		& Type & Setting                                                    
		\\\hline\hline
		\shortstack[l]{I\\\vspace{1em}} & 
		\shortstack[l]{global\\model\\\vspace{0.5em}} & 
		\shortstack[l]{\\$F_{\nu}^{-1}(\cdot)=\eta+\sigma\Phi^{-1}(\cdot)$,\\where $\eta|Z\sim N(\eta_0+\alpha Z, \tau^2)$ and \\$\sigma|Z\sim \mathrm{Gamma}[\{\sigma_0+\beta Z\}^2/\kappa, \kappa/\{\sigma_0+\beta Z\}]$}                 
		\\\hline
		\shortstack[l]{II\\\vspace{1em}} & 
		\shortstack[l]{local\\model \\\vspace{0.5em}} & 
		\shortstack[l]{\\$F_{\nu}^{-1}(\cdot)=\eta+\sigma\Phi^{-1}(\cdot)$,\\where $\eta|Z\sim N(\eta_0+\alpha\sin(\pi Z), \tau^2)$ and \\$\sigma|Z\sim \mathrm{Gamma}[\{\sigma_0+\beta\sin(\pi Z)\}^2/\kappa, \kappa/\{\sigma_0+\beta\sin(\pi Z)\}]$} 
		\\\hline
		\shortstack[l]{III\\\vspace{1em}} & 
		\shortstack[l]{global model\\with random transport\\\vspace{0.5em}} & 
		\shortstack[l]{\\$F_{\nu}^{-1}(\cdot)=T\circ\{\eta+\sigma\Phi^{-1}(\cdot)\}$,\\where $\eta|Z\sim N(\eta_0+\alpha Z, \tau^2)$ and \\$\sigma|Z\sim \mathrm{Gamma}[\{\sigma_0+\beta Z\}^2/\kappa, \kappa/\{\sigma_0+\beta Z\}]$}
		\\\hline
		\shortstack[l]{IV\\\vspace{1em}} & 
		\shortstack[l]{local model\\with random transport\\\vspace{0.5em}} & 
		\shortstack[l]{\\$F_{\nu}^{-1}(\cdot)=T\circ\{\eta+\sigma\Phi^{-1}(\cdot)\}$,\\where $\eta|Z\sim N(\eta_0+\alpha\sin(\pi Z), \tau^2)$ and \\$\sigma|Z\sim \mathrm{Gamma}[\{\sigma_0+\beta\sin(\pi Z)\}^2/\kappa, \kappa/\{\sigma_0+\beta\sin(\pi Z)\}]$} 
		\\\hline
	\etab
	\label{tab:simusetting}
\end{table}

We consider $n=50, 100, 200, 500, 1000$, with $Q=1000$ Monte Carlo runs for each of the four simulation settings. For each $n$, the sample size $N_i$ is independently sampled from the Poisson distribution with parameter $0.25n$. In each Monte Carlo run, predictors $\{Z_i\}_{i=1}^n$ are independently sampled from $U(-1, 1)$. For each $i$, mean and standard deviation of $\nu_i$ are generated conditionally on $Z_i$ as described in Table~\ref{tab:simusetting}, with $\eta_0=0, \sigma_0=3, \alpha=3, \beta=0.5, \tau = 0.5$, and $\kappa = 1$. $N_i$ random observations $\{Y_{ij}\}_{j=1}^{N_i}$ are then independently sampled from $\nu_i$. For the $q$th Monte Carlo run for a given setting, the quality of the estimation is quantified by the integrated squared error (ISE)
\[\mathrm{ISE}_q=\int_{-1}^1d_{\mathcal{W}}^2\{\hat{m}_q(z), m(z)\}dz,\]
where $m(\cdot)$ is the true regression function  and  $\hat{m}_q(z)$ is the fitted regression function. 
The bandwidths for the local REM in the second simulation setting are chosen as $n^{-1/5}$.

\begin{figure}[tb]
	\centering
	\includegraphics[width=\linewidth]{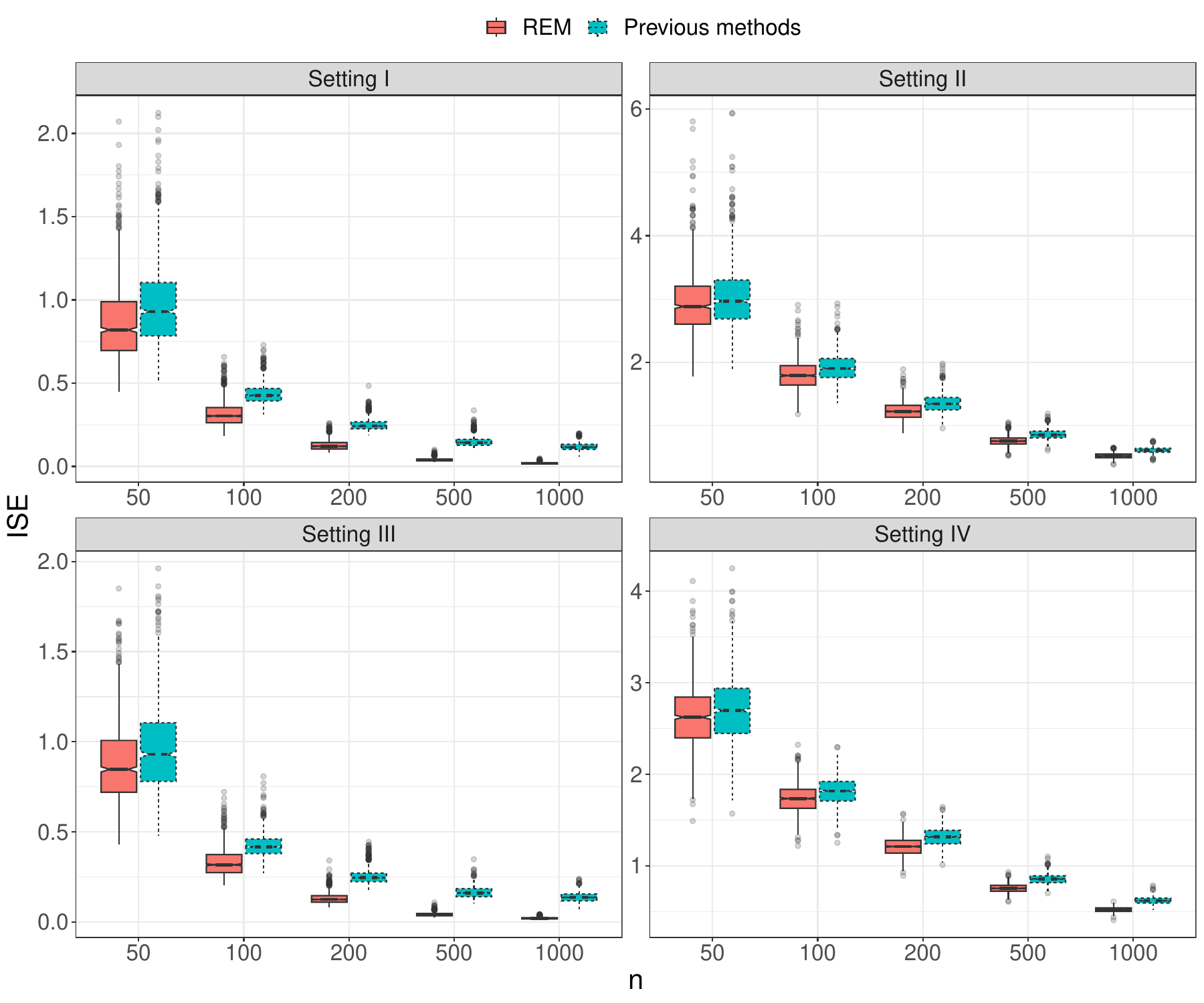}
	\caption{Boxplots of integrated square errors (ISE) for $Q=1000$ Monte Carlo runs and  different simulation settings using the proposed REM (red) and the two-step approaches that were adopted previously (blue).}
	\label{fig:ise}
\end{figure}

We also include comparisons with the previous two-step procedures that included a prior kernel smoothing step  \citep{pana:16, mull:19:3}  to estimate the densities of the $\nu_i$  from the available discrete observations. The integrated squared errors (ISE) for all Monte Carlo runs and different $n$ under the four simulation settings using the proposed methods and the comparison methods are summarized in the boxplots in Figure~\ref{fig:ise}. With increasing sample size, ISE is seen to decrease for all simulation settings, demonstrating the convergence of the proposed methods to the target. Importantly, the proposed REM outperforms the comparison methods under all  simulation settings. Indeed, the preliminary smoothing step required in the previous literature involved a unnecessary tuning parameter selection and fails when the data for some measures are very sparse. 

\section{Data Applications}
\subsection{Cohort-specific BMI distribution for US preschool children}
The Environmental influences on Child Health Outcomes (ECHO) program is an NIH-funded nationwide consortium of multiple cohort studies across the United States designed to investigate the effects of early life exposures on child health and development \citep{gill:18}. The ECHO program combines existing prenatal and pediatric data collected via cohort-specific protocols with a standardized ECHO-wide protocol that was established in 2019 \citep{knap:23}. The de-identified data on participants contributing extant and new data can be accessed through the National Institute of Child Health and Human Development (NICHD) Data and Specimen Hub (DASH); the  version we use here has been made available on  August 31, 2021 \citep{gill:22}. As a multi-cohort study, ECHO brings separate cohorts together so that researchers can access information from a large and diverse population of children followed from the prenatal period through adolescence. 

It is of interest to study the role of demographic factors in child development, measured in terms of body mass index (BMI), calculated as weight in kilograms divided by height in meters squared. We extracted weight and height measurements of preschool children aged approximately 4 years for 17 cohorts from ECHO, along with demographic information for each cohort, aiming to shed light on how the distribution of BMI of preschool children varies across different cohorts in relation to cohort-specific demographic characteristics. The responses specifically are  the cohort-specific distributions of BMI  for 4-year-old children in the respective cohort. Cohort-specific covariates that reflect important demographic characteristics of each cohort include average BMI of mothers, average parental education and proportion of Asians. 

\begin{figure}[p]
 \begin{minipage}{1\textwidth}
     \centering
     \includegraphics[width=0.9\linewidth]{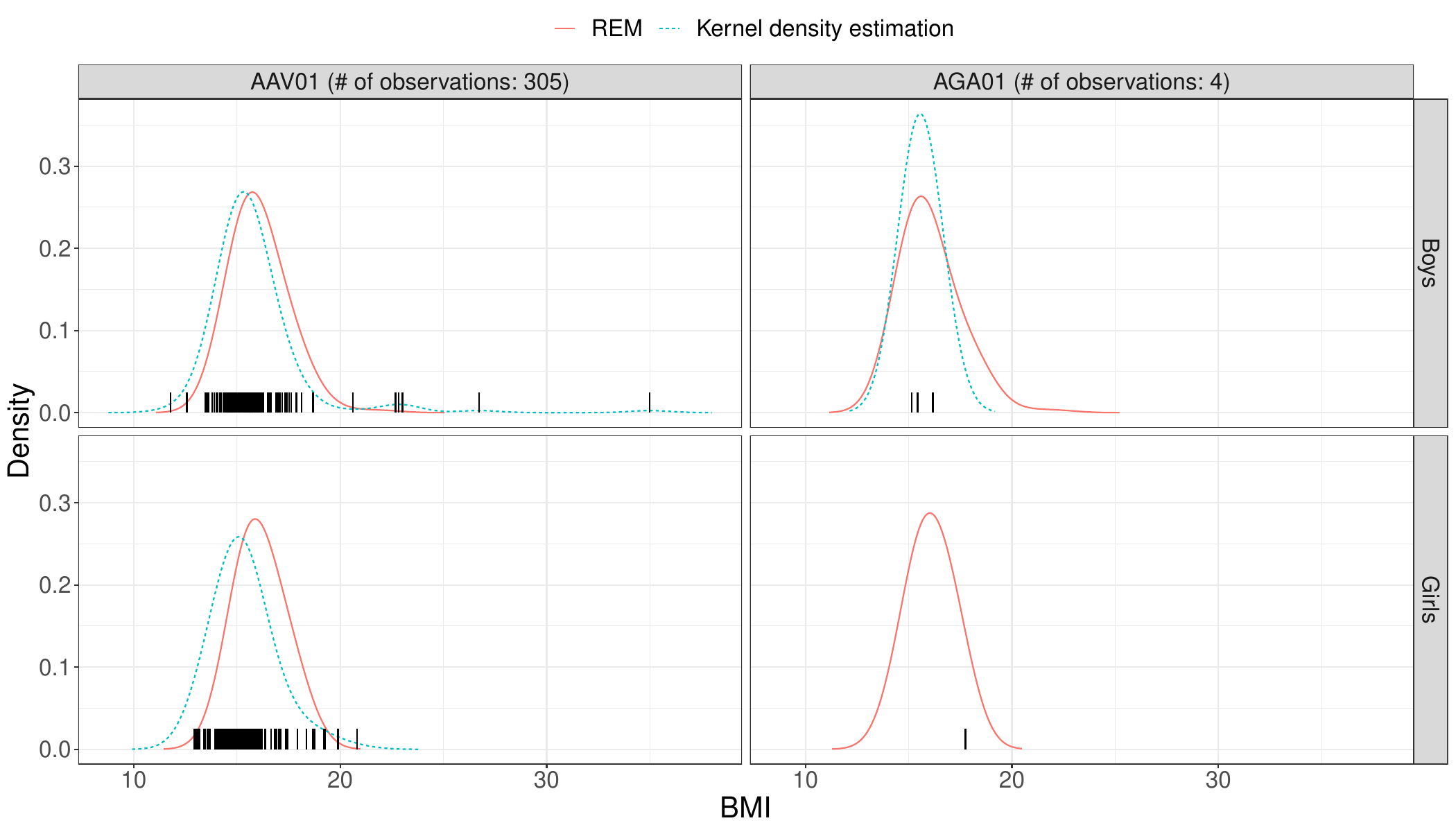}\\
     (a)
   \end{minipage}\hfill
   \begin{minipage}{1\textwidth}
     \centering
     \includegraphics[width=0.9\linewidth]{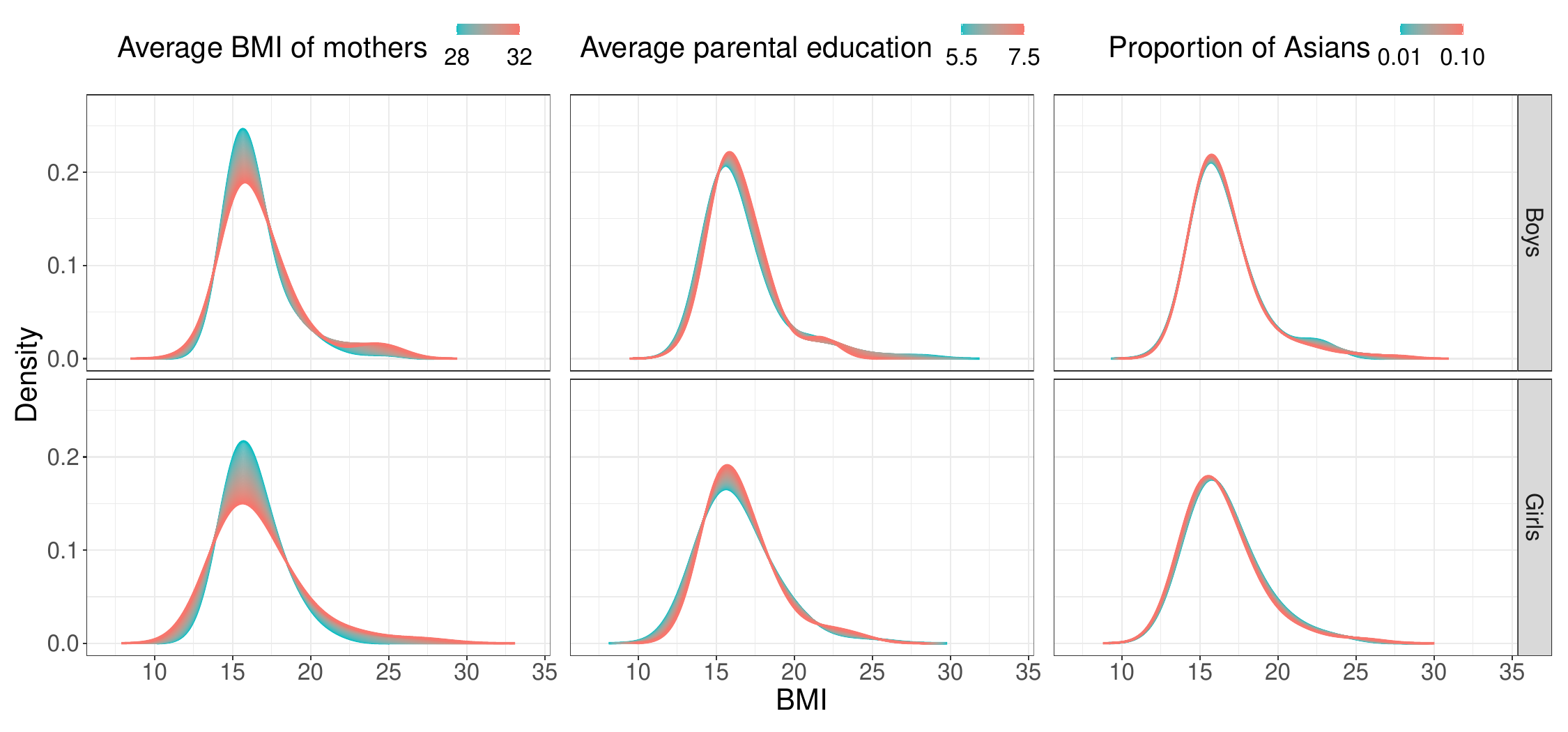}\\
     (b)
    \end{minipage}\\
    \begin{minipage}{\textwidth}
  \caption{(a) Fitted densities of BMI distributions  (solid) of US preschool  boys and girls for AAV01 and AGA01 cohorts, along with direct kernel density estimates (dashed). The corresponding BMI measurements are shown as ticks. (b) Predicted BMI densities of US preschool boys and girls at different predictor levels.}
  \label{fig:echoBmi}
\end{minipage}
\end{figure}

 Due to differences in the cost and accessibility of visits for different cohorts, there is a significant variation in sample size for each cohort. The amount of accessible weight and height measurements for each cohort varies substantially as a result; see Table \ref{tab:echo} for further information. Specifically, there is only one weight and height measurement for girls in AGA01 cohort, while 160 measurements are available for boys in AAV01 cohort. We applied global REM to boys and girls separately. The fitted BMI densities for two cohorts with the most and least weight and height measurements (AAV01, AGA01) are demonstrated in Figure~\ref{fig:echoBmi}(a), along with the corresponding BMI measurements shown as ticks. Although only one measurement is available for girls in the AGA01 cohort, the proposed regression model is able to borrow information from cohorts with more measurements to construct a reasonable density estimate. We also include kernel density estimates in Figure~\ref{fig:echoBmi}(a) for comparison, where the covariate information is not taken into account. Note that kernel density estimation is infeasible for girls in the AGA01 cohort since only one observation is available.
 

 \begin{table}[tb]
	\caption{Number of weight and height measurements for each cohort.}
	\centering
	\btab{lllllllll}
		\hline
		Cohort & AAA01 & AAD01 & AAE01 & AAF01 & AAJ01 & AAU01 & AAV01 & AAW02
		\\\hline
		Boys & 139 & 70 & 77 & 35 & 9 & 10 & 160 & 12 \\
		Girls & 101 & 70 & 81 & 15 & 7 & 8 & 145 & 8 \\
		Total & 240 & 140 & 158 & 50 & 16 & 18 & 305 & 20 \\\hline
	\etab
	\\
	\btab{lllllllll}
		\hline
		AAX04 & AAX06 & ADA01 & AGA01 & AJA02 & AJA03 & AKA01 & AKA02 & ALA01
		\\\hline
		83 & 124 & 8 & 3 & 44 & 6 & 7 & 76 & 134 \\
		67 & 110 & 7 & 1 & 38 & 14 & 3 & 71 & 128 \\
		150 & 234 & 15 & 4 & 82 & 20 & 10 & 147 & 262 \\\hline
	\etab
	\label{tab:echo}
\end{table}

To further investigate the effects of different demographic factors, predicted BMI densities at different predictor levels are shown in Figure~\ref{fig:echoBmi}(b). Separately for  each predictor, the predictor level was varied from the first to the third quantile of the sample, while  the other  two predictors were held fixed at their median level. We observe that the average BMI of mothers is the most influential predictor, where with increasing average BMI of mothers, the BMI distribution for both 4-year-old boys and girls flattens out and becomes less concentrated.  For the predictor average parental education, higher values are associated with heavier tails of the BMI distribution for both 4-year-old boys and girls. For higher average parental education, the BMI distribution of boys also shifts a little to the right. The proportion of Asians in the population  seems to have little effect on the BMI distribution of preschool children.

\begin{figure}[tb]
	\centering
	\includegraphics[width=0.9\linewidth]{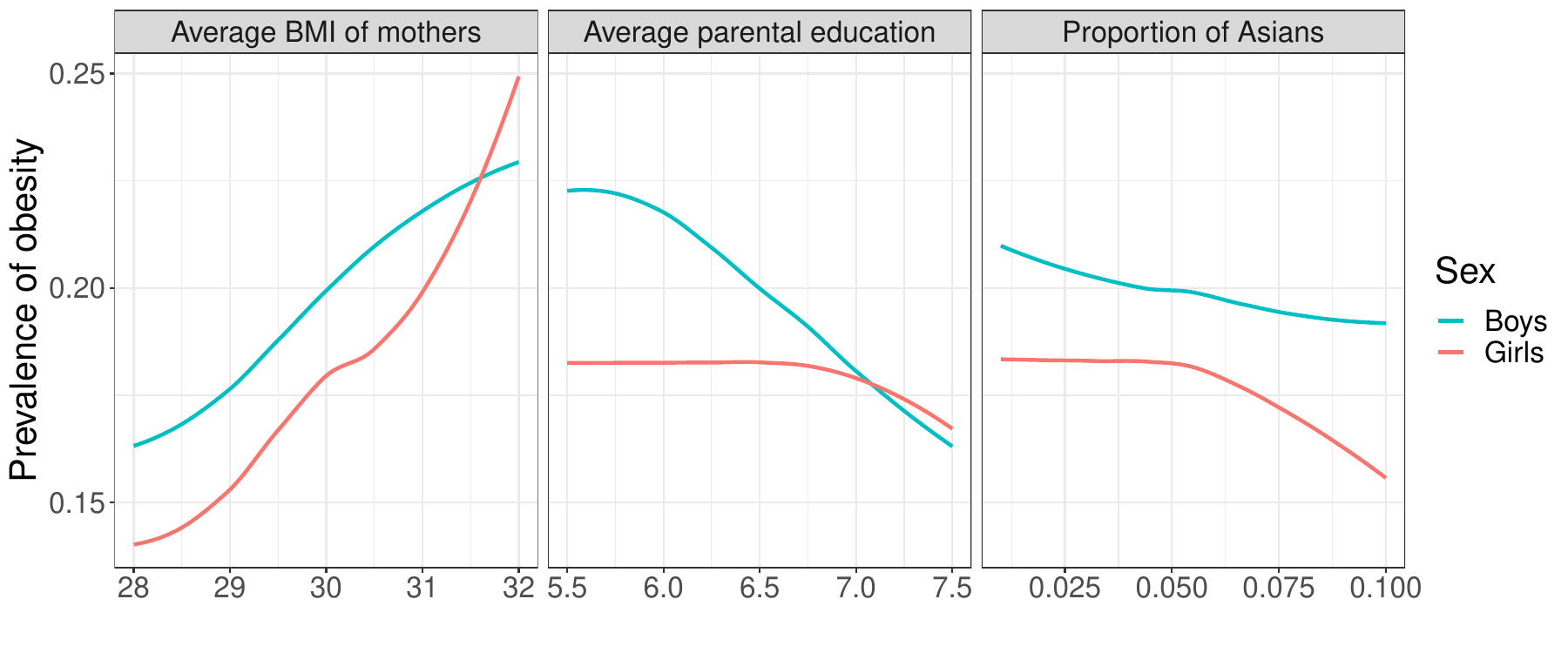}
	\caption{Prevalence of obesity for US preschool boys and girls at different predictor levels.}
	\label{fig:echoObesity}
\end{figure}

Childhood obesity is continuing to rise in the US, and currently about 13.7 million children are considered to be overweight/obese. For the ECHO data, we can compute the probability of obesity from the BMI distribution to investigate demographic disparities in early childhood obesity. Obesity is defined as a BMI at or above the 95th percentile for children of the same age and sex. According to the sex-specific BMI-for-age 2000 CDC Growth Charts, the BMI threshold for a 4-year-old boy to be obese is 17.8, while for girl the value is 18. Based on the predicted BMI distributions (see Figure~\ref{fig:echoBmi}), we calculated the corresponding probabilities of obesity at different predictor levels, with results illustrated in Figure~\ref{fig:echoObesity}. The average BMI of mothers is found to be positively correlated with the prevalence of obesity among four-year-old children, while average parental education and proportion of Asians are negatively associated with obesity. None of these associations 
establishes a causal effect, but these findings are  in agreement with the current literature on obesity,  where parental overweight and low socioeconomic status were found to be strong risk factors of obesity in children \citep{dani:04, vazq:20}, while obesity is less common in Asian children \citep{ande:09}. 

\subsection{Bid time distribution for eBay auction data}
Electronic commerce and in particular eBay, with the prevalence of online auction, is a rich data source for the study of bidding behavior and strategies \citep{wang:08:2, mull:09:1, liu:19}. Here we implemented the proposed regression model to analyze eBay auction data consisting of 93 7-day auctions for an Xbox game console. The data are publicly available at \url{https://www.modelingonlineauctions.com/datasets} \citep{jank:10}. For each auction, the opening bid set by the seller serves as predictor and the time stamps where bids were placed as responses.  We aim to investigate the dependence of bid time distribution on the opening bid. The number of bids for each auction varies from 2 to 75, necessitating the application of the proposed regression model, for which we used local REM. 

\begin{figure}[p]
 \begin{minipage}{1\textwidth}
     \centering
     \includegraphics[width=0.9\linewidth]{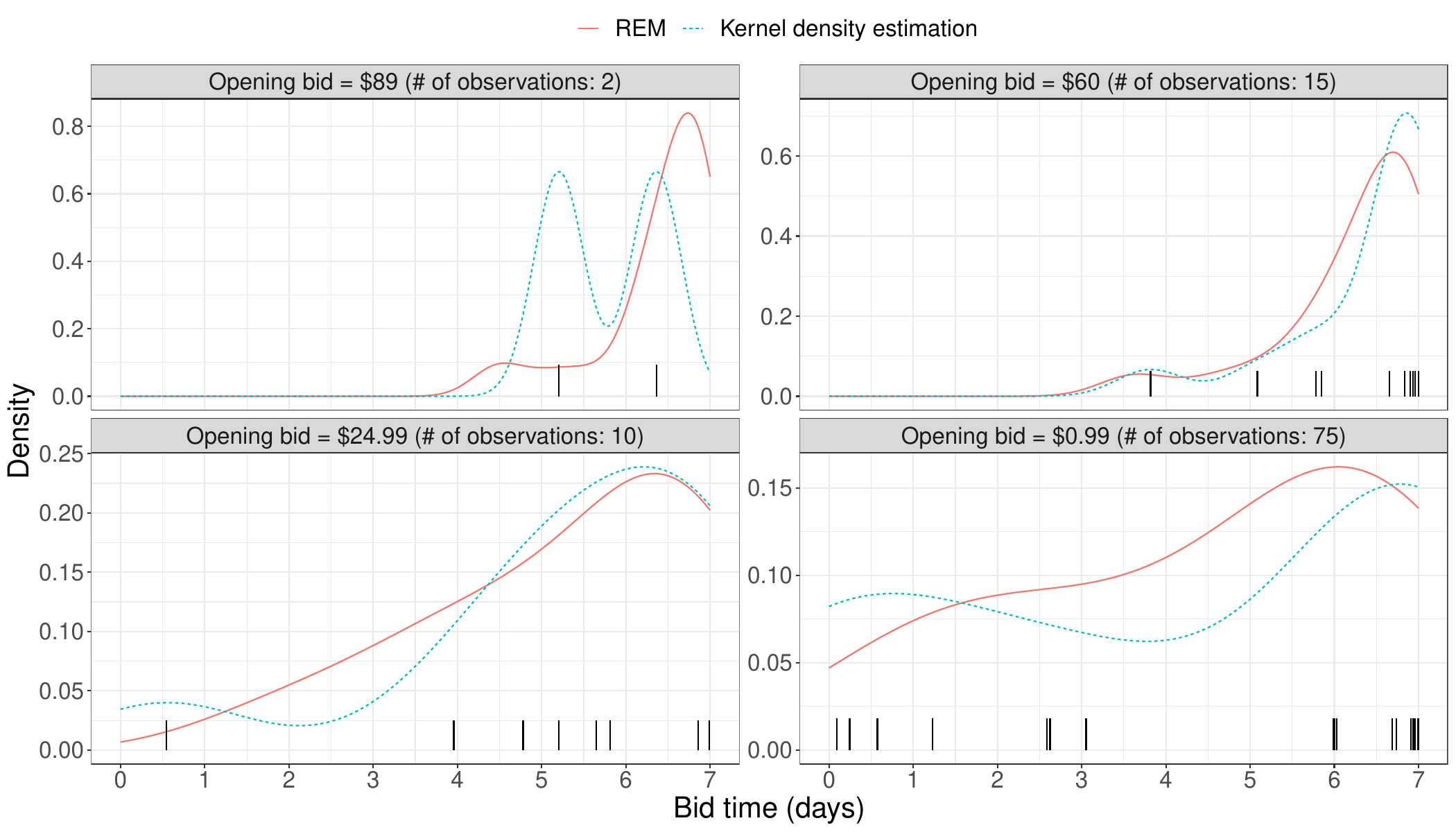}\\
     (a)
   \end{minipage}\hfill
   \begin{minipage}{1\textwidth}
     \centering
     \includegraphics[width=0.9\linewidth]{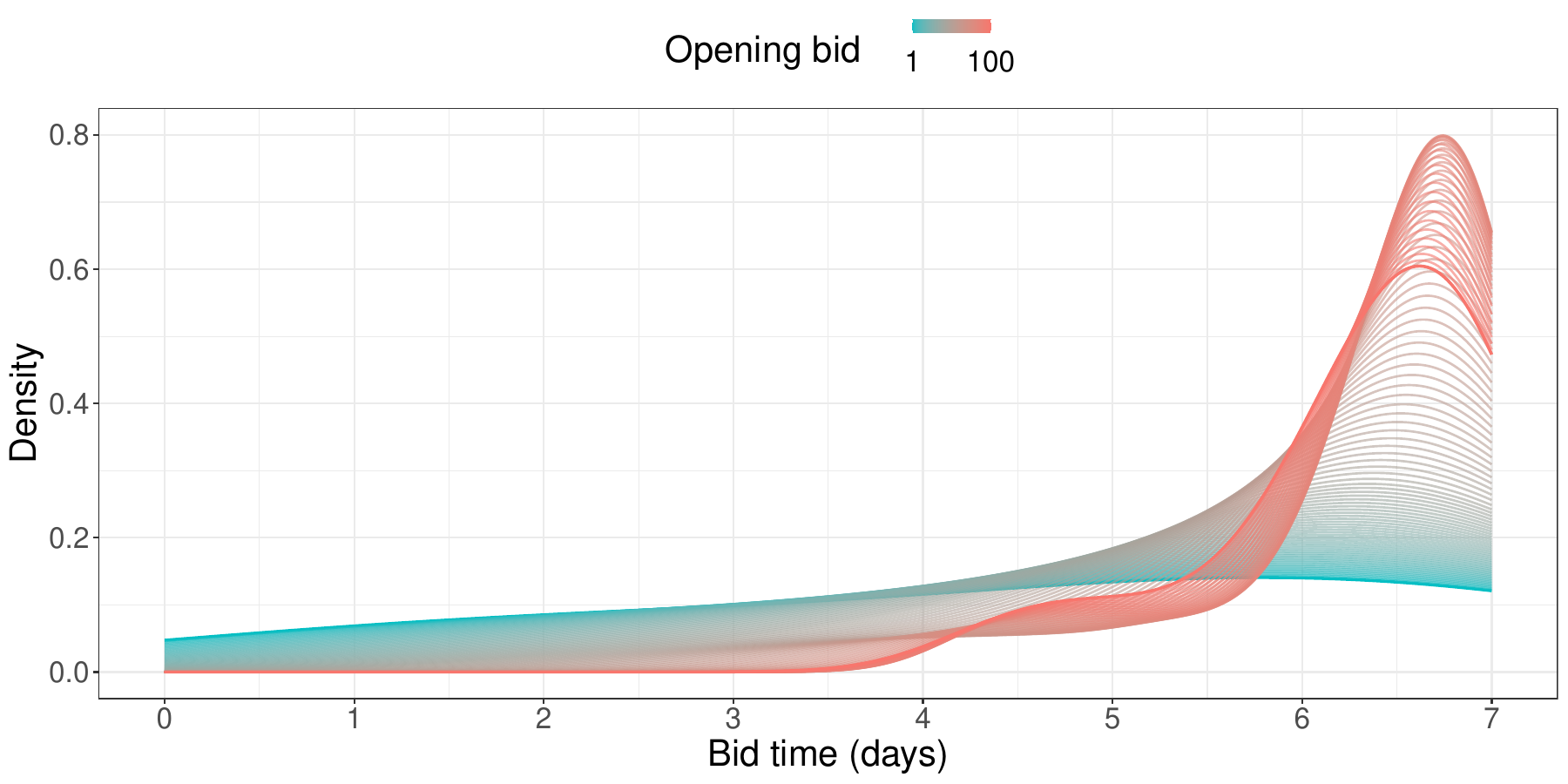}\\
     (b)
    \end{minipage}\\
    \begin{minipage}{\textwidth}
  \caption{(a) Fitted bid time densities (solid) for four different auctions of an Xbox game console with the proposed method,  along with auction-specific kernel density estimates (dashed). The corresponding bid times are shown as ticks. (b) Predicted bid time densities for different opening bids.}
  \label{fig:bidTime}
\end{minipage}
\end{figure}

The fitted bid time densities for four auctions including the two with the most and least bids are illustrated in Figure~\ref{fig:bidTime}(a). The corresponding bid times are shown as ticks for comparison. We find that the proposed regression model is robust to the strong heterogeneity inherent in the auction data, where for example the bid time density for an auction with only two bids (top left panel in Figure~\ref{fig:bidTime}(a)) is reasonably well recovered. In contrast, standard kernel density estimation implemented separately for each auction lacks the ability to borrow strength across the entire sample of distributions and, as expected, fails to reconstruct the underlying bid time density for the auction with only two bids. To further investigate the effect of the opening bid amount on the bid time distribution, we show in Figure~\ref{fig:bidTime}(b) the predicted bid time densities for opening bids varying from 1 to 100 dollars. Bidders tend to bid late in the auction if the opening bid set by the seller is high. In contrast, if the opening bid is low, bid time is more evenly distributed over  the 7-day period. Regardless of the opening bid, bid sniping is prevalent,  where the bidding becomes frenzied near the end of an auction. A lower opening bid is associated with less bid sniping in the sense that bidding is more spread out over the entire time range of the auction.

\section*{Acknowledgements}
\label{sec:akd}
We acknowledge support from NSF grants DMS-20146260, DMS-2310450 and  IOS-2102953.

\appendix
\section*{Appendix A. Assumptions for Theorem \ref{thm:1} and Theorem \ref{thm:2}}
In the following, $f_Z(\cdot)$ and $f_{Z\mid\nu}(\cdot, \mu)$ stand for the marginal density of $Z$ and the conditional density of $Z$ given $\nu=\mu$, respectively, and $\mathcal{T}$ is a closed interval in $\mathbb{R}$ with interior $\mathcal{T}^o$.
\bem[label=(A\arabic*), leftmargin=1cm]
\item The kernel $K(\cdot)$ is a probability density function, symmetric around zero. Furthermore, defining $K_{kj}=\int_{\mathbb{R}}K^k(u)u^jdu$, $|K_{14}|$ and $|K_{26}|$ are both finite.\label{itm:lp0}
\item $f_Z(\cdot)$ and $f_{Z\mid\nu}(\cdot, \mu)$ both exist and are twice continuously differentiable, the latter for all $\mu\in\mathcal{W}$, and $\sup_{z, \mu}|(\partial^2 f_{Z\mid\nu}/\partial z^2)(z, \mu)|<\infty$. Additionally, for any open set $U\subset\mathcal{W}$, $\int_UdF_{\nu\mid Z}(z, \mu)$ is continuous as a function of $z$.\label{itm:lp1}
\item The kernel $K(\cdot)$ is a probability density function, symmetric around zero, and uniformly continuous on $\mathbb{R}$. Furthermore, defining $K_{jk}=\int_{\mathbb{R}}K(u)^ju^kdu$ for $j, k\in\mathbb{N}$, $|K_{14}|$ and $|K_{26}|$ are both finite. The derivative $K'$ exists and is bounded on the support of $K$, i.e., $\sup_{K(x)>0}|K'(x)|<\infty$; additionally, $\int_{\mathbb{R}}x^2|K'(x)|(|x\log|x||)^{1/2}dx<\infty$.\label{itm:lu0}
\item $f_Z(\cdot)$ and $f_{Z\mid\nu}(\cdot, \mu)$ both exist and are continuous on $\mathcal{T}$ and twice continuously differentiable on $\mathcal{T}^o$, the latter for all $\mu\in\mathcal{W}$. The marginal density $f_Z(\cdot)$ is bounded away from zero on $\mathcal{T}$, $\inf_{z\in\mathcal{T}}f_Z(z)>0$. The second-order derivative $f''_Z$ is bounded, $\sup_{z\in\mathcal{T}^o}|f''_Z(z)|<\infty$. The second-order partial derivatives $(\partial^2 f_{Z\mid\nu}/\partial z^2)(\cdot, \mu)$ are uniformly bounded, $\sup_{z\in\mathcal{T}^o, \mu\in\mathcal{W}}|(\partial^2 f_{Z\mid\nu}/\partial z^2)(z, \mu)|<\infty$. Additionally, for any open set $U\subset\mathcal{W}$,  $\int_UdF_{\nu\mid Z}(z, \mu)$ is continuous as a function of $z$;  for any $z\in\mathcal{T},$  $M(\mu, z)=E\{d_{\mathcal{W}}^2(\nu, \mu)|Z=z\}$ is equicontinuous, i.e.,
\[\limsup_{x\to z}\sup_{\mu\in\mathcal{W}}|M(\mu, x)-M(\mu, z)|=0.\]\label{itm:lu1}
\eem

\single
\bibliographystyle{rss}
\bibliography{collection}
	
\end{document}